\documentstyle[11pt]{article}
 
\title{{\bf The Painlev\'e Test of Higher Dimensional KdV Equation}}

\author{{\sc YU} Song-Ju and Takeshi {\sc FUKUYAMA}\\
        \small{\it{Department of Physics, Ritsumeikan University}}\\
        \small{\it{ Kusatsu, Shiga 525 JAPAN }}\\
        {\it e-mails:} \small{fpc30017\,@\,bkc.ritsumei.ac.jp, fukuyama\,@\,bkc.ritsumei.ac.jp}}

\date{}

\begin{document}

\maketitle

\thispagestyle{empty}

\begin{abstract}
We argue the integrability of the generalized KdV(GKdV) equation using the 
Painlev\'e test. For $d( \le 2)$ dimensional space, GKdV equation passes 
the Painlev\'e test but does not for $d \geq 3$ dimensional space. We also 
apply the Ablowitz-Ramani-Segur's conjecture to the GKdV equation in order to 
complement the Painlev\'e test. 
\vskip 20mm

\end{abstract}

\clearpage
%%%%%%%%%%%%%%%%%%%%%%%%%%%%%%%%%%%%%%%%%%%%%%%%%%%%%%%%%%%%%%%%%%%%%%%%%
It is one of the most important challenges in the non-linear integrable  
physics to extend its spatial dimension to two, three, $\cdots$. One step to 
this generalization is the cylindrical and spherical extension of one 
dimensional integrable systems. Toda lattice and KdV equations are 
generalized to cylindrical Toda and cylindrical KdV equations. We know that 
both these generalized systems are also integrable systems. Then it is quite 
natural to ask whether spherical generalization to higher dimensions becomes 
also integrable or not. The purpose of the present letter is to answer to 
this question through the Painlev\'e test using KdV equation.

$d$-dimensional generalization of KdV(GKdV) equation is given by\cite{naka}

\begin{equation}
u_t + 6 u u_x + u_{xxx} + \frac{(d-1)}{2 t} u = 0, \label{eq:gkdv}
\end{equation}
where $d=1$, $2$ and $3$ correspond to KdV, cylindrical KdV and spherical KdV 
equations, respectively.

We expand dependent variable $u(x,t)$ around the movable singular manifold 
$\mu$\cite{wtc,agb}

\begin{eqnarray}
& &u(x,t) \equiv \phi^{-j}_{(x,t)} \sum^{\infty}_{l = 0} u_l(x,t) \phi^l(x,t), \label{eq:uexsm} \\
& &\mu = \{(x,t) \mid \phi(x,t) = 0 \}, \label{eq:sm}
\end{eqnarray}
where $j$($\geq 0$) is integer. Substituting Eq(\ref{eq:uexsm}) into GKdV
equation(\ref{eq:gkdv}), we get the leading term,

\begin{equation}
6 j u_0 \phi^{-2 j - 1} + j (j + 1) (j + 2) \phi^2_x \phi^{-j - 3} = 0. \label{eq:lterm}
\end{equation}

It follows that 

\begin{equation}
j = 2 \label{eq:lead}
\end{equation}
from the condition of the coincidence of $\phi$'s power.

Also we can find resonaces at $l = -1$, $4$, $6$. \hspace{1mm}$l = -1$ 
corresponds to arbitrariness of singularity manifold $\mu(\phi(x,t))$. We can 
determine $u_l$'s by requiring that every power of $\phi$ should satisfy 
Eq.(\ref{eq:gkdv}).

\begin{eqnarray}
\phi^{-5} &;& \hspace{0.2cm} u_0 = -2 \phi^2_x, \label{eq:mfive} \\
\phi^{-4} &;& \hspace{0.2cm} u_1 = 2 \phi_{xx}, \label{eq:mfour} \\
\phi^{-3} &;& \hspace{0.2cm} \phi_t \phi_x + 6 u_2 \phi^2_x - 3 \phi^2_{xx} + 4 \phi_x \phi_{xxx} = 0, \label{eq:mthree} \\
\phi^{-2} &;& \hspace{0.2cm} \frac{(d-1) u_0}{2 t} - u_1 \phi_x + u_{0t} - 6 u_1 u_2 \phi_x - 6 u_0 u_3 \phi_x \nonumber \\
& &+ 6 u_2 u_{0x} + 6 u_1 u_{1x} + 6 u_0 u_{2x} - 3 u_{1x} \phi_{xx} - 3 u_{1xx} \phi_x \nonumber \\
& &- u_1 \phi_{xxx} + u_{0xxx} = 0, \label{eq:mtwo} \\
\phi^{-1} &;& \hspace{0.2cm} \frac{(d-1) u_1}{2 t} + u_{1t} + [6 u_3 u_0 + 6 u_2 u_1 + u_{1xx}]_x = 0, \label{eq:mone} \\
\phi^0 &;& \hspace{0.2cm} \frac{(d-1) u_2}{2 t} + u_3 \phi_t + u_{2t} + 6 u_2 u_3 \phi_x - 6 u_5 \phi^3_x \nonumber \\
& &+ 13 u_3 \phi_{xxx} +6 u_2 u_{2x} + 15 u_{3x} \phi_{xx} + 3 u_{3xx} \phi_x \nonumber \\
& &+ u_{2xxx} = 0, \label{eq:zero} \\
\phi^1 &;& \hspace{0.2cm} \frac{(d-1) u_3}{2 t} + u_{3t} + 6 u^2_3 \phi_x + 18 u_5 \phi_x \phi_{xx} + 6 (u_2 u_3)_x \nonumber \\
& & + 6 u_{5x} \phi^2_x + u_{3xxx} = 0. \label{eq:one} 
\end{eqnarray}

At resonant parts appearing in $\phi^{-1}$ and $\phi^1$terms, the coefficients
 of $u_4$ and $u_6$ are identically vanished. So $u_4$ and $u_6$ remain 
undetermined as is expected. So we put 

\begin{equation}
u_4 = u_6 =0. \label{eq:ufusez}
\end{equation}

If we substitute $u_1$, $u_2$ and $u_3$ determined by the preceding equations 
into it, Eq(\ref{eq:mone}) for $\phi^{-1}$ term is identically satisfied 
irrelevant to $d$. However, the situation is different for $\phi^1$ term. The 
same substitution of $u_l$'s into Eq(\ref{eq:one}) gives

\begin{equation}
\frac{(d-1) (d-2)}{12 t^2 \phi_x} = 0. \label{eq:phionecond}
\end{equation}

Thus only GKdV equations with $d=1$ and $d=2$ pass the Painlev\'e test.

We know that solutions of KdV equation($d=1$) and cylindrical KdV equation
($d=2$) are obtained by uses of Hirota's method\cite{an,anhhc} and of 
the inverse scattering method.\cite{cd1,cd2,cd3}

Therefore we have reassured the well known result. Important is the fact that 
$d \geq 3$ dimensional extension does not pass the Painlev\'e test. Does 
it mean that these higher dimensional extensions are not integrable? We can 
say that the nonlinear system is integrable if it passes the Painlev\'e test. 
But is the converse of this statement true at least in this case? In order to 
answer to this question, we show for $d \geq 3$ that $u$ has a logarithmic 
divergence. 

We have mentioned that we can not have an arbitrary function $u_6$ for 
$(d-1) (d-2) \ne 0$. So instead of Eq.(\ref{eq:uexsm}) we consider the 
expansion\cite{hie}

\begin{equation}
u(x,t) \equiv \phi^{-2}_{(x,t)} \sum^{\infty}_{l = 0,(l \ne 6)} u_l(x,t) \phi^l(x,t) + (u_6 + v_6 \log \phi ) \phi^4. \label{eq:uexsmlog}
\end{equation}
 
Substituting Eq.(\ref{eq:uexsmlog}) into GKdV equation(\ref{eq:gkdv}), we 
obtain Eqs.(\ref{eq:mfive}) $\sim$ (\ref{eq:zero}) and the new equation in 
place of Eq.(\ref{eq:one})

\begin{eqnarray}
\phi^1 &;& \hspace{0.2cm} \frac{(d-1) u_3}{2 t} + u_{3t} + 6 u^2_3 \phi_x + 18 u_5 \phi_x \phi_{xx} + 6 (u_2 u_3)_x \nonumber \\
& & + 6 u_{5x} \phi^2_x + u_{3xxx} + 14 v_6 \phi^3_x= 0. \label{eq:onelog}
\end{eqnarray}

Repeating the same procedure as in the previous case we obtain

\begin{equation}
\frac{(d - 1) (d - 2)}{12 t^2 \phi_x} + 14 v_6 \phi^3_x = 0 \label{eq:phionecondlog}
\end{equation}
and the consistency is restored for $(d - 1) (d - 2) \ne 0$ case. Thus $u(x,t)$
 has a movable logarithmic branch point. This suggests strongly that GKdV 
equations with $d \geq 3$ are not integrable.\cite{mv}

In order to compliment the Painlev\'e test we argue GKdV equation from another 
side concerning with the Ablowitz-Ramani-Segur's(ARS) conjecture.\cite{ars}

ARS-conjecture is summarized as follows. If every nonlinear ordinary 
differential equation obtained by an exact reduction of a nonlinear partial 
differential equation are integrable, then the original partial differential 
equation is also integrable.

We assume $u_2=u_3=0$.Then Eq.(\ref{eq:uexsm}) becomes 

\begin{equation}
u(x,t) = 2 (\log \phi)_{xx}, \label{eq:fft}
\end{equation}
where we have used Eq.(\ref{eq:mfive}) and Eq.(\ref{eq:mfour}).Direct substitution of Eq.(\ref{eq:fft}) into Eq.(\ref{eq:gkdv}) yields the bilinear 
form of GKdV equation

\begin{equation}
(D_t D_x + D^4_x + \frac{d-1}{2 t} \partial_x) \phi \cdot \phi = 0. \label{eq:bgkdv}
\end{equation}

Here $D$ is the Hirota's $D$ operator.\cite{hi}

We expand $\phi$ as follows

\begin{equation}
\phi \equiv 1 + \epsilon \phi^{(1)} + \epsilon^2 \phi^{(2)} + \cdots \label{eq:e}
\end{equation}
and substitute it into GKdV bilinear equation(\ref{eq:bgkdv}). Comparing both 
hand sides of equation order by order of $\epsilon$, we obtain

\begin{eqnarray}
\epsilon &;& (\partial_t \partial_x + \partial^4_x + \frac{d-1}{2 t} \partial_x) \phi^{(1)} = 0, \label{eq:eone} \\
\epsilon^2 &;& 2 (\partial_t \partial_x + \partial^4_x + \frac{d-1}{2 t} \partial_x) \phi^{(2)} = \nonumber \\
& &- (D_t D_x + D^4_x + \frac{d-1}{2 t} \partial_x) \phi^{(1)} \cdot \phi^{(1)}. \label{eq:etwo} 
\end{eqnarray}

We only consider one-soliton solution, i.e, $\phi^{(k)} = 0$, $k \geq 2$ since 
it is suffice to test the integrability. Then Eqs (\ref{eq:eone}) and 
(\ref{eq:etwo}) become 

\begin{equation}
\frac{1}{(\alpha t')^{3/\beta}} \phi^{(1)}_{3z} - \frac{z}{\beta t'} \phi^{(1)}_z + \frac{\gamma}{t'^{\delta}} \phi^{(1)} + \frac{d-1}{2 t'} \phi^{(1)} = 0 \label{eq:eonedash}
\end{equation}
and

\begin{equation}
\frac{3}{(\alpha t')^{3/\beta}} {\phi^{(1)}_{2z}}^2 - \frac{3 z}{\beta t'} {\phi^{(1)}_z}^2 + \frac{3 \gamma}{t'^{\delta}} \phi^{(1)} \phi^{(1)}_z +  \frac{2 (d-1)}{t'} \phi^{(1)} \phi^{(1)}_z = 0.  \label{eq:etwodash}
\end{equation}

Here we have performed independent variables transformation, 

\begin{equation}
z = \frac{x}{(\alpha t)^{1/\beta}}, t' = t  \label{eq:vtf}
\end{equation}
and have assumed that 

\begin{equation}
\phi^{(1)}_{t'} \equiv \frac{\gamma}{t'^{\delta}} \phi^{(1)}. \label{eq:tdepphi}
\end{equation}

Assuming furthermore that the variables can be separated, the solutions to 
Eqs.(\ref{eq:eonedash}) and (\ref{eq:etwodash}) are classified into the 
following five cases.
 
\begin{enumerate}

\item Case of $3/\beta \ne \delta$, $3/\beta \ne 1$ and $\delta \ne 1$

\begin{equation}
\phi^{(1)} = 0. \label{eq:baione}
\end{equation}

\item Case of $3/\beta \ne \delta$, $3/\beta \ne 1$ and $\delta = 1$

\begin{equation}
\phi^{(1)} = 0.
\end{equation}

\item Case of $3/\beta = \delta \ne 1$

\begin{equation}
\phi^{(1)} = 0.
\end{equation}

\item Case of $3/\beta = 1$ and $\delta \ne 1$

\begin{equation}
\phi^{(1)} = 0.
\end{equation}

\item Case of $3/\beta = \delta = 1$

Only this case allows non trivial solution.

Eqs.(\ref{eq:eonedash}), (\ref{eq:etwodash}) become 

\begin{equation}
(3 \gamma + 2 d - 3) {\phi^{(1)}_z}^2 - (3 \gamma + d - 1) \phi^{(1)} \phi^{(1)}_{2z} = 0. \label{eq:baifour}
\end{equation}

Then this case is further divided into subgroup (a) and (b). (a) is 
furthermore classified into sub-subgroup i. and ii..

\begin{enumerate}

\item Case of $d=2$

Eq.(\ref{eq:baifour}) becomes

\begin{equation}
(3 \gamma + 1) ({\phi^{(1)}_z}^2 - \phi^{(1)} \phi^{(1)}_{2z}) = 0. \label{eq:deqtwo}
\end{equation}

\begin{enumerate}

\item Case of $3 \gamma + 1 = 0$

We assume $\phi^{(1)} \equiv T(t) Z(z)$, then $T(t) \propto t^{1/3}$ and Eqs.
(\ref{eq:eonedash}), (\ref{eq:etwodash}) become

\begin{eqnarray}
& &\frac{1}{\alpha} Z_{3 z} - \frac{z}{3} Z_z + \frac{1}{6} Z = 0, \label{eq:deqtwoone} \\
& &\frac{3}{\alpha} Z_{2 z}^2 - z Z_z^2 + Z Z_z = 0. \label{eq:deqtwotwo}
\end{eqnarray}

Here we introduce new dependent variable by $Z_z \equiv - \psi^2$, then we 
obtain from Eqs.(\ref{eq:deqtwoone}) and (\ref{eq:deqtwotwo})

\begin{equation} 
\psi_{zz} = z \psi. \label{eq:airyfunc}
\end{equation}

Therefore we get a special solution of $2$-dimensional case using Airy 
function($Ai(z)$)

\begin{equation}
\phi^{(1)}(z,t) = C t^{- \frac{1}{3}} \int^{\infty}_z dz' Ai(z')^2, \label{eq:ssone}
\end{equation}
where $C$ is a constant.

\item Case of $3 \gamma + 1 \ne 0$

It follows that 

\begin{equation}
{\phi^{(1)}_z}^2 - \phi^{(1)} \phi^{(1)}_{2z} = 0. \label{eq:sstwo}
\end{equation}

We cannot define $\phi^{(1)}$.
\end{enumerate}

\item Case of $d \ne 2$

Substitution of $\phi^{(1)} \equiv T(t) \exp(\lambda(z))$ into Eq.
(\ref{eq:baifour}) yields the equation for $\lambda(z)$

\begin{equation}
\lambda_{2z} - \frac{d-2}{3 \gamma + d - 1} \lambda_z^2 = 0. \label{eq:lambdaeq}
\end{equation}
Therefore we get another special solution for $d (\ne 2)$-dimensional case 

\begin{equation}
\phi^{(1)}(z,t) = C t^{\gamma} (z+z_0)^{\frac{3 \gamma + d - 1}{2 - d}}, \label{eq:sstwotwo}
\end{equation}
where $C$ and $z_0$ are constants.

\end{enumerate}

\end{enumerate}

From Eq.(\ref{eq:sstwotwo}) it is concluded that GKdV equation(\ref{eq:gkdv}) 
with $d \geq 3$ is nonintegrable if $\frac{3 \gamma + d - 1}{2 - d}$ is 
fractional. However, we fix $\gamma$ as $3 \gamma + 1 = 0$, Eq.
(\ref{eq:sstwotwo}) has only a movable singularity.

We did not and could not exhaust the reductions in the sense of ARS conjecture.
 Eq.(\ref{eq:sstwotwo}) of course does not imply that GKdV equations with 
$d \geq 3$ are integrable. As we have mentioned these equations are 
probably nonintegrable. Definite conclusion, however, may be obtained when the 
relationships between the miscellaneous integrability tests are made clear.

\end{document}